\input harvmac
\input epsf
\noblackbox
%%% Paragraphs
\newcount\figno
\figno=0
\def\fig#1#2#3{
\par\begingroup\parindent=0pt\leftskip=1cm\rightskip=1cm\parindent=0pt
\baselineskip=11pt
\global\advance\figno by 1
\midinsert
\epsfxsize=#3
\centerline{\epsfbox{#2}}
\vskip 12pt
\centerline{{\bf Figure \the\figno} #1}\par
\endinsert\endgroup\par}
\def\figlabel#1{\xdef#1{\the\figno}}

\def\pano{\par\noindent}

%%% special math symbols
\def\pmb#1{\setbox0=\hbox{#1}%
 \kern-.025em\copy0\kern-\wd0
 \kern.05em\copy0\kern-\wd0
 \kern-.025em\raise.0433em\box0 }
\font\cmss=cmss10
\font\cmsss=cmss10 at 7pt
\def\rlx{\relax\leavevmode}
\def\Cop{\relax\,\hbox{$\inbar\kern-.3em{\rm C}$}}
\def\Rop{\relax{\rm I\kern-.18em R}}
\def\Nop{\relax{\rm I\kern-.18em N}}
\def\Pop{\relax{\rm I\kern-.18em P}}
\def\Zop{\rlx\leavevmode\ifmmode\mathchoice{\hbox{\cmss Z\kern-.4em Z}}
 {\hbox{\cmss Z\kern-.4em Z}}{\lower.9pt\hbox{\cmsss Z\kern-.36em Z}}
 {\lower1.2pt\hbox{\cmsss Z\kern-.36em Z}}\else{\cmss Z\kern-.4em
 Z}\fi}

%%% misc.

\def\half{{1\over 2}}

\def\ie{{\it i.e.}}
\def\IAprime{IA'}

\def\arrow{\rightarrow}

%%% References

\lref\Adams{J.F.\ Adams, {\it Vector fields on spheres}, Ann.\
of Math.\ {\bf 75}, 603 (1962).}

\lref\Atiyah{M.\ Atiyah, {\it K-theory}, W.A. Benjamin, (1964).}

\lref\BGtypeO{O.\ Bergman, M.R.\ Gaberdiel, {\it Dualities of Type 0  
strings}, JHEP {\bf 9907}, 022 (1999); {\tt hep-th/9906055}.}

\lref\BGH{O.\ Bergman, E.G.\ Gimon, P.\ Ho\v{r}ava, {\it Brane
transfer operations and T-duality of non-BPS states}, JHEP {\bf 9904},
010 (1999); {\tt hep-th/9902160}.}

\lref\BGK{O.\ Bergman, E.G.\ Gimon, B.\ Kol, {\it Strings on
orbifold lines}, JHEP {\bf 0105}, 019 (2001); {\tt hep-th/0102095}.} 

\lref\Bianchi{M.\ Bianchi, {\it A note on toroidal compactifications
of the Type I superstring and other superstring vacuum configurations
with 16 supercharges}, Nucl.\ Phys.\ {\bf B528}, 73 (1998); 
{\tt hep-th/9711201}.}

\lref\BottTu{R.\ Bott, T.\ Tu, {\it Differential forms in algebraic
topology}, Springer, GTM, 1982.} 

\lref\TFB{J.\ de Boer, R.\ Dijkgraaf, K.\ Hori, A.\ Keurentjes,
J.\ Morgan, D.R.\ Morrison, S.\ Sethi, {\it Triples, fluxes, and 
strings}, {\tt hep-th/0103170}.}

\lref\CHL{S.\ Chaudhuri, G.\ Hockney, J.D.\ Lykken, {\it Maximally 
supersymmetric string theories in $D<10$}, Phys.\ Rev.\ Lett.\ 
{\bf 75}, 2264 (1995); {\tt hep-th/9505054}.}

\lref\DP{A. Dabholkar, J. Park, {\it Strings on orientifolds},
Nucl.\ Phys.\ {\bf B477}, 701 (1996); {\tt hep-th/9604178}.}

\lref\TatBog{T.\ Dasgupta, B.\ Stefa\'nski, {\it Non-BPS states and
Heterotic-Type I' duality}, Nucl.\ Phys.\ {\bf B572}, 95 (2000); 
{\tt hep-th/9910217}.}

\lref\EMWone{D.-E.\ Diaconescu, G.\ Moore, E.\ Witten, {\it E8 gauge  
theory, and a derivation of K-theory from M-theory}, 
{\tt hep-th/0005090}.} 

\lref\EMWtwo{D.-E.\ Diaconescu, G.\ Moore, E.\ Witten, {\it A
derivation of K-theory from M-theory}, {\tt hep-th/0005091}.}

\lref\gabstef{M.R.\ Gaberdiel, B.\ Stefa\'nski, {\it Dirichlet
branes on orbifolds}, Nucl.\ Phys.\ {\bf B578}, 58 (2000); 
{\tt hep-th/9910109}.}  

\lref\gutperle{M.\ Gutperle, {\it Non-BPS D-branes and enhanced
symmetry in an asymmetric orbifold}, JHEP {\bf 0008}, 036 (2000);
{\tt hep-th/0007126}.}

\lref\horavawitten{P.\ Ho\v{r}ava, E.\ Witten, {\it Heterotic and Type
I string dynamics from eleven dimensions}, Nucl.\ Phys.\ {\bf B460},
506 (1996); {\tt hep-th/9510209}. }

\lref\hl{L.\ Houart, Y.\ Lozano, {\it Brane descent relations in
M-theory }, Phys.\ Lett.\ {\bf B479}, 299 (2000); 
{\tt hep-th/0001170}.  }

\lref\ikk{K.\ Intriligator, M.\ Kleban, J.\ Kumar, {\it Comments on
unstable branes}, JHEP {\bf 0102}, 023 (2001); {\tt hep-th/0101010}. } 

\lref\Karoubi{M.\ Karoubi, {\it K-theory: an introduction}, Springer 
(1978).} 

\lref\uranga{O.\ Loaiza-Brito, A.M.\ Uranga, {\it The fate of the
type I non-BPS D7-brane}, {\tt hep-th/0104173 }.}

\lref\vafawitten{C.\ Vafa, E.\ Witten, {\it Dual string pairs with N=1 
and N=2 supersymmetry in four dimensions}, Nucl.\ Phys.\ Proc.\
Suppl.\ {\bf 46}, 225 (1996); {\tt hep-th/9507050}.} 

\lref\WittenK{E.\ Witten, {\it D-branes and K-theory}, JHEP {\bf 9812},
019 (1998); {\tt hep-th/9810188}.}

\lref\wittenWVS{E.\ Witten, {\it Toroidal compactification without
vector structure}, JHEP {\bf 9802}, 006 (1998); {\tt hep-th/9712028}}

\lref\senrev{A.\ Sen, {\it Non-BPS states and branes in string
theory}, {\tt hep-th/9904207}.}

\lref\sentwo{A.\ Sen, {\it Stable non-BPS bound states of BPS
D-branes}, JHEP {\bf 9808}, 010 (1998); {\tt hep-th/9805019}.}

\lref\bgone{O.\ Bergman, M.R.\ Gaberdiel, {\it Stable non-BPS
D-particles}, Phys.\ Lett.\ {\bf B441}, 133 (1998);
{\tt hep-th/9806155}.}

\lref\sensix{A.\ Sen, {\it BPS D-branes on non-supersymmetric
cycles}, JHEP {\bf 9812}, 021 (1998); {\tt hep-th/9812031}.}

\lref\bgthree{O.\ Bergman, M.R.\ Gaberdiel, {\it  Non-BPS states in 
Heterotic -- Type IIA duality}, JHEP {\bf 9903}, 013 (1999);
{\tt hep-th/9901014}.}

\lref\senfour{A.\ Sen, {\it $SO(32)$ spinors of Type I and other
solitons on brane-antibrane pair}, JHEP {\bf 9809}, 023 (1998);
{\tt hep-th/9808141}.}

\lref\senfive{A.\ Sen, {\it  Type I D-particle and its interactions},
JHEP {\bf 9810}, 021 (1998); {\tt hep-th/9809111}.}

\lref\frau{M.\ Frau, L.\ Gallot, A.\ Lerda, P.\ Strigazzi,
{\it Stable non-BPS D-branes in Type I string theory},
Nucl.\ Phys.\ {\bf B564}, 60 (2000); {\tt hep-th/9903123}.}

\lref\brudis{I.\ Brunner, J.\ Distler, {\it Torsion D-branes in
nongeometrical phases}, {\tt hep-th/0102018}.}

\lref{\braun}{V.\ Braun, {\it K-Theory torsion}, {\tt hep-th/0005103}.} 

\lref{\gukov}{S.\ Gukov, {\it K-Theory, reality, and orientifolds}, 
Commun.\ Math.\ Phys.\ {\bf 210}, 621 (2000); {\tt hep-th/9901042}.}

%%% Title page
\Title{\vbox{
\hbox{hep-th/0108202}
\hbox{KCL-MTH-01-37}
\hbox{DAMTP-2001-79}}}
{\vbox{\centerline{Non-BPS D-branes and M-theory}}}
\centerline{Matthias R.\ Gaberdiel\footnote{$^\dagger$}{{\tt e-mail: 
mrg@mth.kcl.ac.uk}}$^{,a}$ and 
Sakura Sch\"afer-Nameki\footnote{$^\star$}{{\tt e-mail: 
S.Schafer-Nameki@damtp.cam.ac.uk}}$^{,b}$}
\bigskip
\centerline{\it $^a$Department of Mathematics, King's College London} 
\centerline{\it Strand, London WC2R 2LS, U.K.}
\centerline{\it $^b$Department of Applied Mathematics and Theoretical
Physics}  
\centerline{\it Wilberforce Road, Cambridge CB3 OWA, U.K.}
\smallskip
\vskip2cm
\centerline{\bf Abstract}
\bigskip
\noindent
A dual pair of supersymmetric string theories that involves an 
asymmetric orbifold and an orientifold of Type II is considered.
The D-branes of the orbifold theory (that were recently determined by
Gutperle) are all non-BPS and do not carry any conserved gauge
charges. It is shown that they carry non-trivial K-theory charges, and
that they can be understood in terms of branes wrapping certain
homology classes of the M-theory compactification. Using the adiabatic
argument, dual partners of some of these non-BPS D-branes are
proposed. The relations between these dual states are found to be in
agreement with the M-theory description of the D-branes.   
\bigskip

\Date{08/2001}

\newsec{Introduction}
\pano
Recently much progress has been made in the study of non-BPS solitonic
states in string theory \refs{\senrev}. A number of stable non-BPS
D-branes have been constructed explicitly, most notably for certain 
orbifold theories \refs{\sentwo,\bgone,\sensix,\bgthree,\gutperle},
and Type I (and Type IA) \refs{\senfour,\senfive,\frau,\BGH,\TatBog}. 
All of these stable non-BPS D-branes can be elegantly characterized in 
terms of K-theory \refs{\WittenK}.  

Non-BPS D-branes play a crucial role for our understanding of the
strong-weak coupling dualities of certain supersymmetric string
theories. While the dualities are most easily tested on the BPS
spectrum of the theory, the duality map actually has to relate the
whole spectrum of the two theories to one another. In particular, it
is therefore interesting to understand what happens to some of the 
non-BPS states under this map. In some examples it was possible to
identify the image of certain perturbative non-BPS states of one
theory with non-BPS D-brane states of the dual theory 
\refs{\senfour,\bgthree,\TatBog}. In all of these cases, a crucial
ingredient for the identification of these non-BPS states was the
fact that they were the lightest states carrying a conserved
gauge charge. 

In this paper we consider a dual pair of supersymmetric string
theories that relates an asymmetric orbifold to an orientifold 
of Type II. The D-branes of the orbifold theory were recently
determined by Gutperle \refs{\gutperle}. All of these D-branes are in
fact non-BPS and do not carry any conserved gauge charges. We  
confirm that these non-BPS D-branes are indeed stable by determining
the corresponding K-theory groups which turn out to be $\Zop_2$ in
each case. (The situation is therefore similar to the case studied
in \refs{\brudis}, see also \refs{\braun}.)
We then explain how these K-theory groups can be thought to
arise from the cohomology of the corresponding M-theory
compactification.\footnote{$^\ddagger$}{Our argument is similar in
spirit to the analysis of \refs{\uranga}.} The relation between
K-theory classes of R-R fields  and cohomology in M-theory has
recently received much attention  \refs{\EMWone,\EMWtwo,\TFB}. In
particular, it was shown in \refs{\EMWone,\EMWtwo} that the partition
function for the R-R $p$-form fields of Type IIA (that are classified
by K-theory) agrees with the partition function for the $p$-form
fields in M-theory (that are classified by a certain subset of
cohomology). Some direct identifications between K-theory and
cohomology classes were also found in \refs{\TFB}.   

The duality of the two string theories can be understood to originate
from the duality of the corresponding theories before orbifolding or
orientifolding, using the adiabatic argument of \refs{\vafawitten}. 
Some of the non-BPS D-branes of the orbifold theory have a direct
interpretation in terms of brane anti-brane pairs in the original
theory. We can therefore use the known duality relations for the
theories before orbifolding and orientifolding to make a proposal 
for the duals of these non-BPS D-branes. These proposals can then be 
checked against the description of the non-BPS D-branes in terms of
branes wrapping homology cycles of the M-theory compactification. We
also make some speculations about the duals of some of the other
non-BPS D-branes.  
\smallskip

The paper is organised as follows. In section~2 we describe the
different theories as well as their D-brane spectra and duality
relations in detail. The latter are confirmed in section~3 by
comparing the mass formulae for the BPS states in the different
theories. In section~4 we explain how the D-branes can be understood
in terms of M-branes wrapping homology cycles of the underlying
M-theory compactification. In section~5 we make some proposals for the
duals of certain non-BPS D-branes of the orbifold theory and perform
various consistency checks. Section~6 contains some conclusions and
open questions. We have included an appendix in which we give the
details of the computation of the K-theory classes for the Type II
orbifold theories.

\newsec{The duality relations and the D-brane spectrum}
\pano
In this paper we want to consider the duality between the
orbifold of Type IIB  
\eqn\orbifold{ \hbox{IIB on} \quad S^1 / (-1)^{F_L} \sigma_{\half}\,,}
and the orientifold 
\eqn\orientifold{ \hbox{IIB on} \quad S^1 / \Omega \sigma_{\half}\,.}
Here $\sigma_{\half}$ denotes the half-shift along an $S^1$ (which we
shall take to lie in the $x^9$ direction). The orientifold theory is
sometimes referred to as Type
$\tilde{\hbox{I}}$.\footnote{$^\star$}{In nine dimensions the moduli
space of orientifold compactifications with sixteen supercharges has
two components \refs{\TFB}: the usual Type I vacuum for which both
orientifold planes carry the same R-R charge, and the Type
$\tilde{\hbox{I}}$ theory for which the two R-R charges are opposite
\refs{\DP,\wittenWVS}. In the latter case the overall R-R charge
vanishes and there is no need to introduce D8-branes, thus leading to
a trivial gauge group.} The duality between the orbifold and 
orientifold theory \refs{\DP,\Bianchi} can be derived from the
self-duality of Type IIB using the  adiabatic argument of
\refs{\vafawitten}.  

We shall also consider the duality in eight dimensions obtained by
compactifying these theories on an additional $S^1$ in the
8-direction. If we T-dualise both theories along $x^8$ we obtain IIA
orbifold and orientifold theories, respectively,
\eqn\theories{
\hbox{IIA on}\quad S^1_8\times S^1_9/(-1)^{F_L} \sigma_{\half}^9
\qquad \hbox{and} \qquad 
\hbox{IIA on}\quad S^1_8\times S^1_9/\Omega I_8 \sigma_{\half}^9 \,,}
where $I_8$ denotes the reflection in the $x^8$ direction. We shall
sometimes refer to the orientifold theory as \IAprime. (This theory is
not the same as what is usually called $\widetilde{\hbox{IA}}$, which
is the T-dual of Type $\tilde{\hbox{I}}$ along $x^9$.) The two IIA
theories are S-dual as well. This can be seen from the point of view
of M-theory \refs{\DP} by compactifying M-theory on a Klein 
bottle times a circle, $(S^1_8 \times S^1_9)/\Zop_2 \times S^1_{10}$, 
where $\Zop_2$ acts as $I_8\sigma_{\half}^9$ together with a sign
change in the three-form potential, $S_{C^3}$. From the point of view
of the orientifold theory the coupling constant is proportional to
$R_{10}^{3\over 2}$, while the coupling constant of the orbifold
theory is proportional to $R_8^{3\over 2}$. The two theories are
therefore related by an `8-10' flip. The relations between the
various theories are sketched in figure~1. 

\fig{Duality relations in 8-dimensions.}{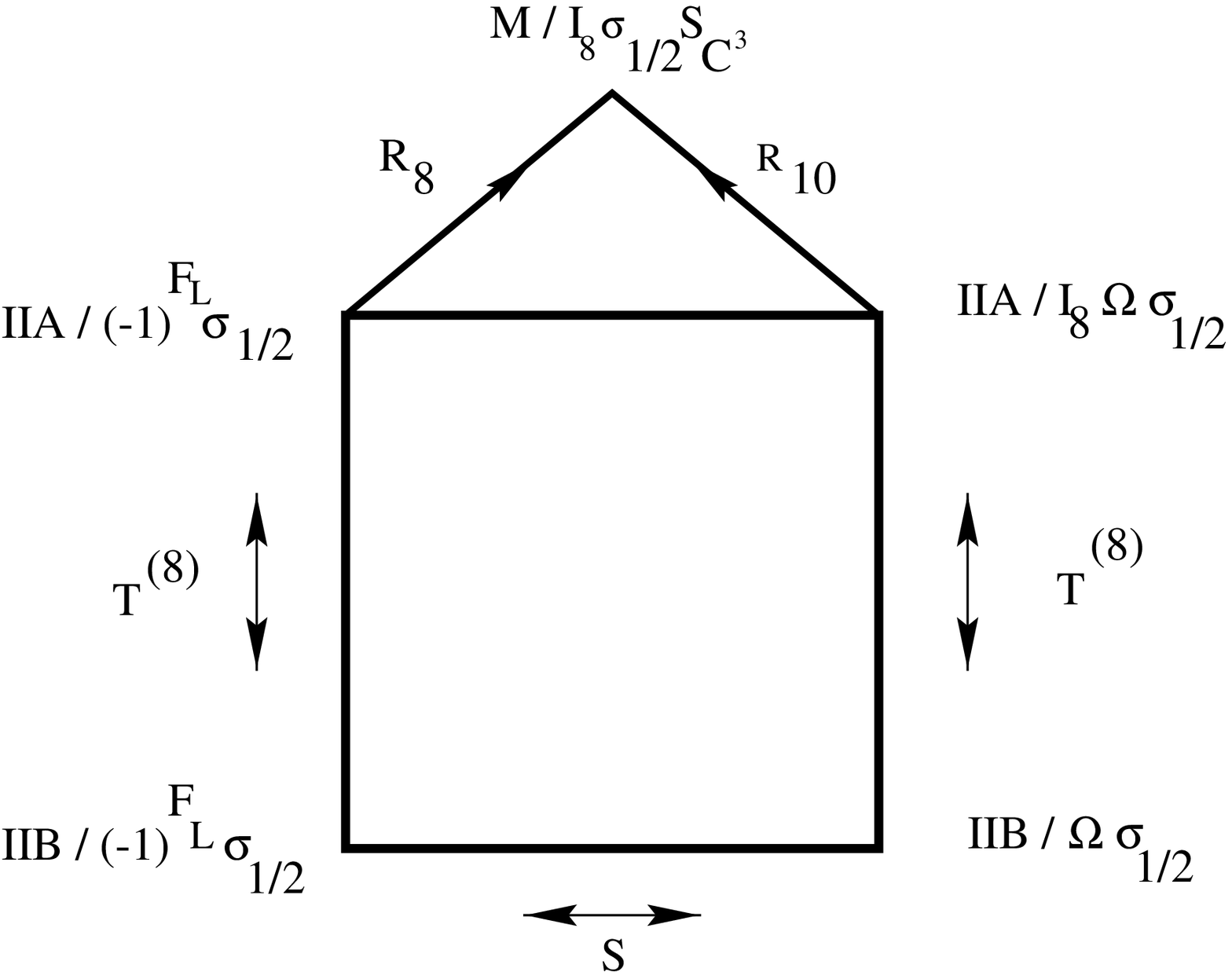}{4.2truein}

\noindent More specifically, the relations between the moduli of the
two IIB theories are given as 
\eqn\dualityIIB{\eqalign{ g_{IIB} & = 4\, g_{\tilde{I}}^{-1} \cr 
R^i_{IIB}  & = \sqrt{2}\, g_{\tilde{{I}}}^{-\half} R^i_{\tilde{{I}}}
\cr 
G_{IIB}  & =  2 \,g_{\tilde{{I}}}^{-1} G_{\tilde{{I}}}\,,}}
where $R_i$ is any of the radii (including $i=9$). Here and in the
following we shall set $\alpha'=1$. Together with the T-duality
relations 
\eqn\Tduality{\eqalign{ 
g_{IA'} & =  {g_{\tilde{{I}}}\over R_{\tilde{{I}}}^8} \cr
R_{IA'}^8 & = {1\over R_{\tilde{{I}}}^{8}} \cr
R^i_{IA'} & = R^i_{\tilde{{I}}} \qquad \hbox{for $i\not= 8$}\cr
G_{IA'} & = G_{\tilde{{I}}}}\quad\quad
\eqalign{ 
g_{IIA} & =  {g_{IIB}\over R_{IIB}^8} \cr
R_{IIA}^8 & = {1\over R_{IIB}^{8}} \cr
R^i_{IIA} & = R^i_{IIB} \qquad \hbox{for $i\not= 8$}\cr
G_{IIA} & = G_{IIB}\,,}
}
this then implies the relations for the IIA moduli
\eqn\dualityIIA{\eqalign{ 
g_{IIA} & = 2 \sqrt{2}\, ( g_{\IAprime})^{-\half} 
(R_{\IAprime}^8)^{3\over 2}
\cr  
R^8_{IIA}  & = {1\over \sqrt{2}} (g_{\IAprime})^\half 
(R^8_{\IAprime})^{\half}\cr 
R^i_{IIA}  & = \sqrt{2} (g_{\IAprime})^{-\half} R^i_{\IAprime} 
(R^8_{\IAprime})^\half \qquad \hbox{for $i\not= 8$} \cr
G_{IIA}  & = 2 (g_{\IAprime})^{-1} R^8_{\IAprime} G_{\IAprime}\,.}}
These relations can be re-derived in terms of M-theory by setting 
\eqn\Mtheory{
R_8=g_{IIA}^{2/3}= 2 { R^8_{\IAprime} \over g_{\IAprime}^{1/3}  }
\qquad \hbox{and}\qquad 
R_{10}=\half g_{\IAprime}^{2/3}= { R^8_{IIA} \over g_{IIA}^{1/3}  }\,,}
together with 
\eqn\Mtheorynine{
R_9 = {R^9_{IIA}\over g^{1/3}_{IIA}} 
= {R^9_{\IAprime}\over g_{\IAprime}^{1/3}}\,.}
Here, the radii without suffix are measured in M-theory. The formulae
differ from those given by Ho\v{r}ava and Witten
\refs{\horavawitten} by some factors of two, reflecting the fact that
the moduli are effectively describing the double cover of the Klein
bottle.

\subsec{The D-brane spectrum}

The orbifold theories have $(0,16)$ supersymmetry, while the
supersymmetry of the orientifold theories is $(8,8)$. The latter
theories therefore have BPS D-branes, while all the D-brane states of
the orbifold theories are necessarily non-BPS. For the following it
will be useful to summarise the D-brane spectra of the different
theories.  

For the case of the IIB orbifold theory in nine dimensions the
D-branes (and their boundary states) have been determined in
\refs{\gutperle}. There are two kinds of branes that are
distinguished by the boundary condition in the compact
9-direction. Following the convention introduced in \refs{\gabstef} we
denote these D-branes as D($r$,0) or D($r$,1), where $r+1$ is the
number of Neumann directions transverse to $x^9$, and the former
branes are Dirichlet with respect to $x^9$, while the latter have a
Neumann boundary condition. It follows from the analysis of
\refs{\gutperle} that both kinds of branes exist provided that $r$ 
is even (odd) in IIA (IIB). Furthermore, for given such $r$, the two
branes D($r$,0) and D($r$,1) can decay into one another depending on
the size of the radius in the compact $x^9$ direction. 

In the spirit of \refs{\WittenK} the charges of these D-branes are
classified by K-theory. Orbifold theories involving the action of
$(-1)^{F_L}$ are described by the Hopkins groups $K_{\pm}$
\refs{\WittenK}, and the K-theory group that classifies D-brane
charges in our context is therefore  
\eqn\kpm{
K_{\pm}(S^{8-r} \times S^{2,0}, S^{2,0})\,.}
Here $r+1$ is the number of Neumann directions transverse to $x^9$. 
In writing \kpm\ we have used the standard notation  
$\Rop^{p+q}\simeq \Rop^{p,q}$ to indicate that the 
(geometrical part of the) $\Zop_2$-orbifold generator acts 
on the first $p$ coordinates as $-1$, and we have denoted by $S^{p,q}$
the $p+q-1$ sphere in $\Rop^{p,q}$; we have also used the short hand
notation $S^p=S^{0,p+1}$. On the circle $S^{2,0}$ the action of
$\Zop_2$ corresponds then precisely to the half-shift
$\sigma_{\half}$ (that accompanies $(-1)^{F_L}$ in the action of the
orbifold generator). We have computed the groups \kpm\ in appendix
A. For the IIB  orbifold we obtain
\eqn\KIIB{
K_{\pm}(S^{8-r} \times S^{2,0}, S^{2,0}) = \left\{
\eqalign{\Zop_2 \qquad &\hbox{if $r$ is odd,}\cr
           0 \quad\quad\;   & \hbox{if $r$ is even.}}\right.}
This is in agreement with the D-brane spectrum that was found by
Gutperle in \refs{\gutperle}.\footnote{$^\dagger$}{Since the two
branes D($r$,0) and D($r$,1) can decay into one another, they define 
the same K-theory class.} For the IIA orbifold we obtain similarly 
\eqn\KIIA{
K^{-1}_{\pm}(S^{8-r} \times S^{2,0}, S^{2,0}) = \left\{
\eqalign{\Zop_2  \qquad &\hbox{if $r$ is even,} \cr
           0 \quad\quad\;    &\hbox{if $r$ is odd.}}\right.}
These K-theory calculations imply in particular that a single non-BPS 
D-brane is (topologically) stable, whereas an even number of such
D-branes can decay into the vacuum. The $\Zop_2$ nature of the charge
can be understood from the fact that the D$(r,0)$ branes may be
described in terms of brane anti-brane pairs in the theory before 
orbifolding \refs{\gutperle}.
\vskip4pt

The D-branes of the IIB orientifold theory in nine dimensions were
determined in \refs{\gukov,\BGH} by computing the K-theory groups
$\widetilde{KSC}(X)$; the results are summarised in table~1. 
\vskip4pt
$$
\def\tbntry#1{\vbox to 23 pt{\vfill \hbox{#1}\vfill }}
\hbox{\vrule 
      \vbox{\hrule 
            \hbox{\vrule
                  \hbox to 70 pt{
                  \hfill\tbntry{$D(r,s)$: $r=$}\hfill }
                  \vrule 
                  \hbox to 20 pt{
                  \hfill\tbntry{$-1$}\hfill }
                  \vrule 
                  \hbox to 20 pt{
                  \hfill\tbntry{$0$}\hfill }
		  \vrule 
                  \hbox to 20 pt{
                  \hfill\tbntry{$1$}\hfill }
                  \vrule 
                  \hbox to 20 pt{
                  \hfill\tbntry{$2$}\hfill }
                  \vrule 
                  \hbox to 20 pt{
                  \hfill\tbntry{$3$}\hfill }
                  \vrule 
                  \hbox to 20 pt{
                  \hfill\tbntry{$4$}\hfill }
                  \vrule 
                  \hbox to 20 pt{
                  \hfill\tbntry{$5$}\hfill }
                  \vrule 
                  \hbox to 20 pt{
                  \hfill\tbntry{$6$}\hfill }
		  \vrule 
                  \hbox to 20 pt{
                  \hfill\tbntry{$7$}\hfill }
                  \vrule 
                  \hbox to 20 pt{
                  \hfill\tbntry{$8$}\hfill }
                  \vrule 
                 }
            \hrule  
            \hbox{\vrule
                  \hbox to 70 pt{
                  \hfill\tbntry{$\widetilde{KSC}(S^{8-r})$}\hfill }
                  \vrule  
                  \hbox to 20 pt{
                  \hfill\tbntry{$\Zop_2$}\hfill }
                  \vrule 
                  \hbox to 20 pt{
                  \hfill\tbntry{$\Zop$}\hfill }
                  \vrule 
                  \hbox to 20 pt{
                  \hfill\tbntry{$\Zop$}\hfill }
                  \vrule 
                  \hbox to 20 pt{
                  \hfill\tbntry{$0$}\hfill }
                  \vrule 
                  \hbox to 20 pt{
                  \hfill\tbntry{$\Zop_2$}\hfill }
                  \vrule 
                  \hbox to 20 pt{
                  \hfill\tbntry{$\Zop$}\hfill }
                  \vrule 
                  \hbox to 20 pt{
                  \hfill\tbntry{$\Zop$}\hfill } 
		  \vrule 
                  \hbox to 20 pt{
                  \hfill\tbntry{$0$}\hfill } 
		  \vrule 
                  \hbox to 20 pt{
                  \hfill\tbntry{$\Zop_2$}\hfill }
                  \vrule 
                  \hbox to 20 pt{
                  \hfill\tbntry{$\Zop$}\hfill }
                  \vrule 
                 }
            \hrule 
         }
     }
$$
\centerline{
\hbox{{\bf Table 1:} {\it D-branes in the IIB orientifold in nine
            dimensions.}}} 
\vskip 8pt

\noindent Upon compactifying on an additional circle and using
T-duality (along $x^8$), the result for the IIA orientifold theory (in
eight dimensions) can be derived from the above.

\newsec{Comparison of BPS states}
\pano
In order to check the duality relations \dualityIIB\ we shall next
compare the mass formulae for the BPS states of the two IIB theories.
We shall deal with the different classes of states in turn.

\noindent {\bf Massless states.} In the NS-NS sector of the
orbifold theory we have the massless bosonic states (with zero
momentum) corresponding to the graviton, the $B_{\mu\nu}$ field and
the dilaton; in the orientifold theory, the graviton and dilaton come
from the NS-NS sector, while the $B_{\mu\nu}$ field arises in the R-R
sector. Again, these are massless (and do not carry any momentum).
\medskip

\noindent {\bf $9$-momentum states.} In the untwisted NS-NS sector
of the orbifold theory we have bosonic states that have even momentum, 
\eqn\one{
(p_L,p_R) = \left( {2m\over R^9_{IIB}}, {2m \over R^9_{IIB}}
\right)\,.} 
Their mass in the orbifold theory is $M_{IIB}={2 |m|\over R^9_{IIB}}$, 
and in Type $\tilde{\hbox{I}}$ the mass is 
\eqn\two{
M_{\tilde{{I}}} = \sqrt{2} g_{\tilde{{I}}}^{-\half} 
{2 |m| \over \sqrt{2} g_{\tilde{{I}}}^{-\half} R^9_{\tilde{{I}}} } 
= {2 |m| \over R^9_{\tilde{{I}}}\,. }}
Thus these states are (closed string) momentum states in Type
$\tilde{\hbox{I}}$ whose momentum is again even. In addition, there
are bosonic states with odd momentum: in the orbifold theory these
arise in the untwisted R-R sector (and therefore have $(-1)^{F_L}$
eigenvalue $-1$), and in the orientifold theory their eigenvalue under
$\Omega$ is $\Omega=-1$.  
\medskip

\noindent {\bf $9$-winding states.} The lightest bosonic state with
$9$-winding arises in the twisted NS-NS sector, and it is
characterised by 
\eqn\three{
\left(p_L,p_R\right) = \left({n R^9_{IIB}\over 2}, 
- {n R^9_{IIB}\over 2} \right) \,,}
where $n\in \Zop$. The IIB mass is $M_{IIB}={|n| R^9_{IIB} \over 2}$,
and in terms of Type $\tilde{\hbox{I}}$ this is 
\eqn\four{
M_{\tilde{{I}}} =  \sqrt{2} g_{\tilde{{I}}}^{-\half} 
{|n| \sqrt{2} g_{\tilde{{I}}}^{-\half} R^9_{\tilde{{I}}} \over 2} 
= { g_{\tilde{{I}}}^{-1} |n| R^9_{\tilde{{I}}} }\,.}
This therefore represents a BPS D(0,1) brane in Type
$\tilde{\hbox{I}}$, {\it i.e.} a D1-brane that wraps the $S^1$ in the
9-direction. This D-brane is invariant under the orientifold
projection (as is also suggested by the K-theory calculation of
table~1). Indeed, on the oscillator exponential of the boundary state, 
both $\Omega$ and $\sigma_{\half}$ act trivially; for the case of a
wrapped D1-brane, the ground state involves a sum over all $9$-winding
numbers, together with an integral over the transverse
momenta. $\Omega$ maps $w_9\mapsto - w_9$, and therefore leaves the
sum invariant (provided the Wilson line is either zero or $\half$),
and $\sigma_{\half}$ acts trivially on winding
states.\footnote{$^\ddagger$}{A priori, there is an ambiguity in
defining the $\Zop_2$ shift action, that differs in the way the shift
operator acts on winding states \refs{\vafawitten}. If the action of 
$\sigma_{\half}$ in  $\tilde{\hbox{I}}$ was non-trivial on winding
states, the boundary state would not be invariant, and the BPS
spectrum of the two theories would not agree.}  

There are $64$ bosonic states of this type (together with $64$
fermionic states that arise in the twisted NS-R sector) in the
orbifold theory, and they form a short multiplet (of dimension $128$)
of the supersymmetry algebra. This agrees precisely with the
degeneracy of the D1-brane BPS state. 
\medskip

\noindent {\bf $8$-momentum states.} The analysis is fairly
analogous to the case of $9$-momentum: the states without momentum and
winding in the $9$-direction all come from the untwisted NS-NS sector,
and they can have arbitrary integer $8$-momentum. Using the same
formula as before, these states correspond to (closed string) states
in $\tilde{\hbox{I}}$ that have arbitrary integer $8$-momentum (and
are invariant under $\Omega$). 
\medskip

\noindent {\bf $8$-winding states.} The case of winding in the
$8$-direction is more interesting. Again, the states without momentum
and winding in the $9$-direction all come from the untwisted NS-NS
sector in the orbifold theory. Their momentum is now given by
\eqn\five{
\left(p_L,p_R\right) = \left(n R^8_{IIB}, - n R^8_{IIB} \right)\,, }
and thus the mass is $M_{IIB}=|n| R^8_{IIB} $. In terms of
Type $\tilde{\hbox{I}}$ this is 
\eqn\six{
M_{\tilde{{I}}} =\sqrt{2} g_{\tilde{{I}}}^{-\half} 
|n| \sqrt{2} g_{\tilde{{I}}}^{-\half} R^8_{\tilde{{I}}}
= 2g_{\tilde{{I}}}^{-1} |n| R^8_{\tilde{{I}}}\,.}
This corresponds to {\it two} D1-branes that wrap the
$8$-direction. This is necessary in order to have an orientifold
invariant combination: the two D1-branes have to sit at the opposite
points of the $x^9$-circle in order to be invariant under
$\sigma_{\half}$. In fact, this is one of the BPS D-branes that was  
discussed in \refs{\BGH}.

There is one interesting lesson that can be drawn from this analysis
that will prove useful later on. In the theory before orbifolding
or orientifolding, S-duality exchanges the D1-brane with the
fundamental string. One may have therefore thought that the
combination of two D1-branes in Type $\tilde{\hbox{I}}$ theory should  
correspond, under S-duality, to the superposition
\eqn\pert{ 
|w_8=+1,x_9=0\rangle + |w_8=+1,x_9=\pi R^9_{IIB}\rangle 
}
in the orbifold theory. (If we take these states to be the (lowest)
GSO-invariant states in the NS-NS (NS-R) sector, these are indeed
invariant under the orbifold projection.) In terms of a
momentum-winding basis, this superposition would then be 
\eqn\pertu{ 
\sum_{m\in\Zop} \left|w_8=+1,p_9={2 m \over R^9_{IIB}}\right\rangle
\,.} 
However, this naive derivation does {\it not} agree with what we have
found based on the analysis of the BPS masses: from that point of
view, only the state with $p_9=0$ arises. We shall use the analogy
with this situation to identify the dual of a non-BPS D-brane later
on.

\subsec{The relation to IIA and M-theory}

Let us briefly comment on what these different states correspond to in
the two IIA theories and in M-theory. A state with winding and
momentum along $x^8$ and $x^9$ in the IIB orbifold theory has mass
\eqn\IIBmass{M_{IIB}^2 
= {n_9^2 \over R_9^2 } + m^2 R_9^2 
+ {k^2 \over R_8^2 } + n_{10}^2 R_{8}^2 \,.}
Under the duality map, this becomes
\eqn\IIAmass{\eqalign{
M_{\tilde{I}}^2 &= { n_9^2 \over R_9^2 } 
+ 4 { m^2 R_9^2 \over g^2} 
+ {k^2 \over R_8^2 } 
+ 4 { n_{10}^2 R_8^2 \over g^2}\cr
M_{IIA}^2 & =  { n_9^2 \over R_9^2 } 
 + m^2 R_9^2 + k^2 R_{8}^2 + { n_{10}^2 \over R_8^2}\cr
M_{\IAprime}^2 & = { n_9^2 \over R_9^2 } 
+ 4 {m^2 R_8^2 R_9^2 \over  g^2 } 
+ k^2 R_8^2 + 4 {n_{10}^2 \over  g^2} \,,}}
where we have written all masses in terms of the moduli of the
respective theories. The BPS states are then identified as follows: 
the IIB 9-momentum states (with even momentum) are mapped to IIA
9-momentum states (with even momentum), which are 9-momentum states
(with even momentum) in \IAprime. A 9-winding state with half-integer 
winding (coming from the twisted sector) is mapped to a single
\IAprime\ D2-brane wrapping the 8- and 9-directions. 8-momentum states
in IIB map to 8-winding states in IIA and 8-winding states in
\IAprime. On the other hand, 8-winding states in IIB have
\IAprime-mass 
\eqn\IAprimemass{
M_{\IAprime}={2 n\over  g} \,,
}
and therefore correspond to {\it two} D0-branes; this is again
required in order to obtain an orientifold invariant configuration.

All of these states can be thought of as arising from M-theory where
they correspond to states with mass 
\eqn\massM{M_M^2
= {n_9^2 \over R_9^2 } + m^2 R_8^2 R_9^2  
+ k^2 R_8^2 R_{10}^2  + { n_{10}^2 \over  R_{10}^2 }\,.} 
The relevant states are KK-momentum states with momentum along 9 and
10, and membrane winding states, respectively. In the untwisted sector
of the orbifold theory, all quantum numbers are integers, but in the
twisted sector we also have states for which the 9-winding number $m$
is half-integer. From the point of view of M-theory, these states will
therefore define new kinds of {\it twisted membrane} states. Given our 
limited understanding of M-theory orbifolds, we do not know how to
derive the existence of these states directly in terms of
M-theory. (In particular, they do not seem to be necessary for the
cancellation of gravitational anomalies as in \refs{\horavawitten};
the situation is therefore similar to what was found in
\refs{\BGtypeO}.)  However, given that the relevant states in the
orbifold theory are BPS, these states must be present in the M-theory
spectrum.

\newsec{K-theory charges from M-theory cohomology}
\pano
Recently it has been proposed that K-theory fluxes in Type IIA string 
theory may be related to certain M-theory cohomology classes
\refs{\EMWone,\EMWtwo,\TFB}. Before proceeding, we want to explore
a related question for the case at hand, namely whether the
K-theoretic D-brane charges of the IIA orbifold theory can be
understood in terms of M-theory cohomology. This is of particular
interest in our case since the D-brane charges are all pure torsion.

Because of Poincar\'e duality (see (4.3) below for the relation in our
case) the corresponding D-branes should be described in terms of
M-branes wrapping suitable homology cycles of the M-theory
compactification. Since all the D-branes of the IIA orbifold are
non-BPS this amounts to identifying the M-theory lift of these non-BPS 
D-brane states. Non-BPS states in M-theory have been discussed 
before in \refs{\hl,\ikk}, where they were constructed out of M-brane
anti-M-brane pairs by tachyon condensation, and in \refs{\uranga}
where they were obtained by lifting non-BPS configurations in a IIA
orientifold theory to 11-dimensions; our analysis here will be similar
in spirit to the latter approach.

For the theory in question M-theory is compactified on a Klein
bottle (times the 10-circle that does not play a role in the 
following since it is common to both M-theory and the IIA orbifold). 
Since the Klein bottle ${\bf K}$ is not orientable there
are two kinds of (co)homologies: normal integral (co)homology and
`twisted' (co)homology which takes coefficients in
$\widehat{\Zop}$. Here $\widehat{\Zop}$ is the $\Zop_2$-module
where $\Zop_2$ acts by the non-trivial representation (for details on
twisted (co)homologies see for example \refs{\BottTu}). The normal
integral homology of ${\bf K}$ is given by   
\eqn\KBhoZ{H_{n}(\bf{K}, \Zop) = \left\{ 
\eqalign{ \Zop\quad\,\;  & \qquad    n=0 \cr
	 \Zop \oplus \Zop_2  & \qquad  n=1 \cr
	 0  \quad\,\, & \qquad   n\geq 2\,, }\right.}
and the corresponding integral cohomology is
\eqn\KBcohoZ{
H^{n}(\bf{K}, \Zop) = \left\{ 
\eqalign{\;\Zop\,\,  & \qquad    n=0  \cr
	 \;\Zop\,\, &\qquad  n=1 \cr
	 \;\Zop_2 &\qquad  n=2 \cr
	 \;0\;\;  & \qquad n\geq 3\,. }\right.}
The `twisted' (co)homology 
$\widehat{H}_*^{(*)}({\bf K},\widehat{\Zop})$ can be  
obtained from a twisted version of Poincar\'e
duality\footnote{$^\star$}{This notion of `twisting' has nothing to do
with the `twisted' sectors of the orbifold.}
\eqn\twistedpoin{
\widehat{H}_n({\bf K}, \widehat{\Zop})= H^{d-n}({\bf K}, \Zop)
\qquad \hbox{and}\qquad 
\widehat{H}^n({\bf K}, \widehat{\Zop})= H_{d-n}({\bf K}, \Zop)\,,
}
where, in our case, $d=2$. 

Under the action of the M-theory orbifold, the 3-form changes sign,
and therefore the membrane reverses its orientation. This implies that
the membrane can only wrap on `twisted' homology cycles 
$\widehat{H}_n({\bf K}, \widehat{\Zop})$. It follows 
from \twistedpoin\ and \KBcohoZ\ that there are two $\Zop$-cycles for
$n=1$ and $n=2$, and a $\Zop_2$ cycle for $n=0$. Wrapping the membrane
around the 2-cycle gives rise to a fundamental string with 9-winding
in the orbifold theory, whereas the membrane that is wrapped around
the 1-cycle produces a fundamental string state without
9-winding. Both of these states are indeed BPS (and carry $\Zop$
charge). On the other hand, the $\Zop_2$ 0-cycle corresponds to the
non-BPS D(2,0) brane of the IIA orbifold theory. This brane can be
thought of as coming from a D2-brane anti-brane pair in the theory
before orbifolding (where the two branes sit at opposite points of the
9-circle). Naively this would seem to lift to a M2-brane anti-brane
pair; however, the above description in terms of a twisted homology
class suggests that the configuration actually comes from a single
M2-brane that `wraps' this twisted cycle.

The other extended object in M-theory is the M5-brane that is
unaffected by the orbifold action, and that should therefore only wrap
around untwisted cycles $H_{n}(\bf{K}, \Zop)$. It follows from \KBhoZ\
that there are two $\Zop$-cycles for $n=0$ and $n=1$, as well as a
$\Zop_2$ cycle for $n=1$. If we `wrap' the M5-brane around the
0-cycle, we obtain the NS 5-brane in the orbifold theory, whilst
wrapping the M5-brane around the $\Zop$ 1-cycle gives rise to a NS
5-brane that wraps the 9-direction. (This is an allowed configuration
since the NS 5-brane is invariant under $(-1)^{F_L}$.) On the other
hand, the $\Zop_2$ 1-cycle corresponds to the non-BPS D(4,0) brane of
the orbifold theory. Again this brane can be thought of as coming from
a D4 anti-D4-brane pair in the theory before orbifolding which naively
lifts to M5 anti-M5-brane pair in M-theory. However, the above
homology analysis suggests again that one can think of this as a
single M5-brane wrapping the $\Zop_2$ homology 1-cycle.

\newsec{Identifying non-BPS states}
\pano
In this section we shall attempt to identify the non-BPS D-brane
states of the IIB orbifold theory with non-BPS states of the dual Type 
$\tilde{\hbox{I}}$ theory. We shall also explain how our proposals tie
in with the description of the T-dual non-BPS states of the IIA
orbifold in terms of M-theory.

As we have reviewed in section~2, there are two types of D-branes
states in the IIB orbifold theory that can decay into one another
depending on the radius of the circle in the 9-direction
\refs{\gutperle} 
\eqn\decay{
(2r-1,0) \longleftrightarrow (2r-1,1) \,.}
The branes of the form D($2r-1,0$) can be thought to consist of a 
D($2r-1,0$) brane of Type IIB at $x^9=a$ together with a D($2r-1,0$)
anti-brane at $x^9=a+\pi R^9$. This is an orbifold invariant
configuration since $(-1)^{F_L}$ maps a brane to an anti-brane. The
boundary state of this non-BPS D-brane only involves the untwisted
sector as should be expected from this interpretation. On the other
hand, the boundary state of the D($2r-1,1$) brane has non-trivial
components in the untwisted NS-NS as well as the twisted R-R sector;
its interpretation in terms of Type IIB branes is therefore not clear.

\subsec{The D(3,0) brane}

There is one non-BPS D-brane of the orbifold theory that can be quite
easily identified with a non-BPS state of Type $\tilde{\hbox{I}}$:
this is the D(3,0) brane that has an interpretation in terms of a
brane anti-brane pair of the Type IIB theory. We know that in the
theory before orbifolding and orientifolding, the D3-brane of Type IIB
is self-dual. Thus we should expect that the D(3,0) brane of the
orbifold theory corresponds to a D-brane in Type $\tilde{\hbox{I}}$
that can be made out of a D3-brane at $x^9=a$ together with an
anti-D3-brane at $x^9=a+\pi R^9$. The corresponding boundary state
describes in fact precisely the D(3,0) brane of Type
$\tilde{\hbox{I}}$ that was constructed in \refs{\BGH}. Thus we have
identified 
\eqn\iden{\eqalign{
\hbox{IIB-orbifold} & \hskip20pt \hbox{Type $\tilde{\hbox{I}}$} \cr
\hbox{D}(3,0) & \longleftrightarrow  \hbox{D}(3,0)\,.}}
Both of these D-branes are $\Zop_2$ charged, and the reason is the
same in both cases: if we have two such branes, we can move the
D3-brane of one combination to come close to the anti-D3-brane of the
other, and they annihilate.
\smallskip

Suppose the D(3,0) does not wrap the 8-direction. Under T-duality
(along $x^8$), the D(3,0) brane then becomes a D(4,0) brane in the IIA
orbifold theory (that wraps the 8-direction of the orbifold theory,
\ie\ the $10$ direction of M-theory). As we have explained before,
this brane lifts in M-theory to the M5-brane wrapping the $\Zop_2$
1-cycle of the Klein bottle (as well as the $10$ direction). Under the
`8-10 flip' this then becomes a D4-brane in the \IAprime\ theory
(since the M5-brane wraps along the $10$ direction), and therefore,
under T-duality, a D3-brane in $\tilde{\hbox{I}}$. This agrees with
\iden.  

On the other hand, if the D(3,0) does wrap the 8-direction, then under
T-duality (along $x^8$) the D(3,0) brane becomes a D(2,0) brane in the
IIA orbifold (that does not wrap the $10$ direction). This brane lifts
to the M2-brane in M-theory that corresponds to the twisted 0-cycle of
the Klein bottle. Under the `8-10 flip' this then becomes again a
D2-brane in \IAprime, and thus a D3-brane in $\tilde{\hbox{I}}$. This
also agrees with \iden.

\subsec{The D(1,0) brane}

Next let us consider the D(1,0) brane of the IIB orbifold theory. As
we have explained before, this brane can be thought to consist of a
D1-brane at $x^9=0$, together with an anti-D1-brane at $x^9=\pi R^9$. 
Under S-duality, one should therefore expect that this state
corresponds to something like  
\eqn\ansatz{ 
|w_8=+1,x_9=0\rangle + |w_8=-1,x_9=\pi R^9\rangle }
in Type $\tilde{\hbox{I}}$. Here we have assumed that the D(1,0) brane
wraps along the $x^8$ direction.

The above argument is fairly analogous to the one we put forward for
the BPS $8$-winding states. In that case we saw that the actual dual
state was not \ansatz, but rather the lowest momentum component; so we
propose now that the dual of the D(1,0) brane of the orbifold theory
is the superposition 
\eqn\actual{
|w_8=+1,p_9=0 \rangle + \epsilon |w_8=-1,p_9=0\rangle \,.}
Here $|w_8,p_9\rangle$ describes the momentum and winding of the $128$  
bosonic states that are the lowest GSO-invariant states in the NS-NS
and R-R sector, and similarly for the fermions. In order for the above
superposition to be invariant under the orientifold projection, we
have to choose $\epsilon$ to be the eigenvalue of the corresponding
state under the action of $\Omega$, {\it i.e.} $\epsilon=+1$ if the
ground state describes the graviton and dilaton in the NS-NS sector or
the $B_{\mu\nu}$ state in the R-R sector, and $\epsilon=-1$ otherwise.  
There are 64 states for which $\Omega=\epsilon=+1$, and 64 states for
which $\Omega=\epsilon=-1$. Together with the 128 fermions these
states form a long (non-BPS) multiplet of the supersymmetry algebra. 
(It is clear that they form one long multiplet rather than two short
multiplets since they are not BPS; indeed, under the duality map their
mass in the orbifold theory is that of a D-brane, and we know that
none of the D-branes of the orbifold theory are BPS.)

One might wonder whether the states in \actual\ are actually stable,
or whether they can decay to states with $w_8=p_9=0$. At $w_8=p_9=0$
there are only 64 physical boson states since only the states with
$\Omega=+1$ survive the orbifold projection. These 64 states (together
with their fermions) form a short multiplet of the supersymmetry
algebra and these states are indeed BPS. Thus it would seem that the
states with $\epsilon=+1$ can decay to $w_8=p_9=0$, but that this is
not possible for the states with $\epsilon=-1$. However, since
multiplets must decay as whole multiplets this means that the
whole long multiplet cannot decay in this way. 

We thus propose that the dual of the non-BPS D(1,0) brane of the IIB
orbifold theory is dual to the state in \actual,   
\eqn\identone{\eqalign{
\hbox{IIB-orbifold} & \hskip20pt \hbox{Type $\tilde{\hbox{I}}$} \cr
D(1,0) & \longleftrightarrow  w^i=\pm 1 \,,}}
where $w^i$ is the winding number in the (compact) direction along
which the D(1,0) brane wraps.
\smallskip

As before, we can check whether this proposal makes sense in the
T-dual IIA picture. Under T-duality, the non-BPS D(1,0) brane (that
wraps $x^8$) maps to the non-BPS D(0,0) brane of the IIA orbifold. The 
D(0,0) brane can be thought of as a D0-brane anti-brane pair of the
theory before orbifolding (where the two branes are at opposite points
of the 9-circle). In M-theory these correspond to a combination of
states that have (positive and negative) 10-momentum. Under the `8-10'
flip we therefore obtain the state in \IAprime
\eqn\actualTdual{ |p_8=1, p_9=0 \rangle  
+  \epsilon |p_8=-1,p_9=0\rangle\,. }  
This is precisely the T-dual (along $x^8$) of \actual.

Similarly, the T-dual of the D(1,0) brane (that wraps $x^7$, say, but
not $x^8$) is the non-BPS D(2,0) brane of the IIA orbifold (that wraps
$x^7$ and $x^8$, \ie\ the 10-direction of M-theory). In M-theory this
lifts to the M2-brane wrapping the $7$ and $10$ direction (as well as
the twisted 0-cycle of the Klein bottle). Under the `8-10 flip' this 
then becomes a fundamental string state in \IAprime\ that has winding
along the 7-direction. This is again in agreement with \identone.

\subsec{A speculation}

As is explained in \refs{\gutperle} the D(3,0) brane of the orbifold 
theory can decay, for sufficiently small $R^9_{IIB}$, to a D(3,1)
non-BPS D-brane. On the other hand, the dual D(3,0) brane of Type
$\tilde{\hbox{I}}$ can decay, for sufficiently small $R^9_{\tilde{I}}$, 
into the non-BPS D$(3,1)$ brane of Type $\tilde{\hbox{I}}$.
It follows from \dualityIIB\ that $R^9_{\tilde{I}}$ is proportional to 
$R^9_{IIB}$, and therefore that the regimes of stability are at least
qualitatively related to one another. This then suggests that we can
also identify
\eqn\ident{\eqalign{
\hbox{IIB-orbifold} & \hskip20pt \hbox{Type $\tilde{\hbox{I}}$} \cr
\hbox{D}(3,1) & \longleftrightarrow  \hbox{D}(3,1)\,.}}
However, it should be clear that this argument is somewhat unreliable
since the actual relation between the moduli is 
$R^9_{IIB}=\sqrt{2} g_{\tilde{I}}^{-\half} R^9_{\tilde{I}}$, and thus 
$R^9_{IIB}$ can also become large if $g_{\tilde{I}}$ becomes small
(for arbitrary $R^9_{\tilde{I}}$). Furthermore, in the regime in
which the M-theory description (that interpolates between the two dual 
string theories) is valid we have $R_8,R_9,R_{10}\gg 1$, and therefore  
both $R^9_{IIB}$ and $R^9_{\tilde{I}}$ are large; in particular, this
suggests that in the M-theory regime, the D$(r,1)$ brane may always
be unstable to decay into the D$(r,0)$ brane. If this is the case we
cannot simply identify the dual of the D$(r,1)$ brane by keeping track
(as we  vary the moduli to go from weak to strong coupling) of the
lightest state with certain properties.

\newsec{Conclusions and outlook}
\pano
In this paper we have analysed the duality between the asymmetric
orbifold IIB$/(-1)^{F_L} \sigma_{\half}$ and the orientifold 
IIB$/\Omega \sigma_{\half}$, as well as their T-dual IIA theories. We
have shown that the non-BPS D-branes of the orbifold theory that had
been constructed by Gutperle \refs{\gutperle} carry $\Zop_2$ K-theory
charge (but do not carry any conserved gauge charge). We have also
explained how these D-branes can be understood in terms of homology
cycles of M-theory. Among other things, this sheds some light on
non-BPS states in M-theory; this is of interest since the tachyon
condensation arguments that underlie for example \refs{\hl} are at
present not very well understood. 

Using the fact that the duality can be understood to originate from a
duality of the theories before orbifolding and orientifolding, we have
made a proposal for the duals of the non-BPS D-branes of the orbifold
theory. We have also shown that these proposals are in agreement with
their M-theory interpretation. Since the non-BPS states do not carry
any conserved gauge charges, this identification of non-BPS states
goes beyond what had been achieved before.  
\smallskip

It would be interesting to perform a similar analysis for the case
of the CHL string \refs{\CHL} which corresponds to M-theory on a
M\"obius strip. It would also be illuminating to understand how the
(non-BPS) D-branes of the Type IA theory can be understood in terms of 
M-theory homology cycles, and in particular, how the gauge charges of
the D-string \refs{\TatBog} arise from this point of view. Finally, it
may be interesting to study fluxes in these models along the lines of 
\refs{\TFB}.

\vskip 1cm

\centerline{{\bf Acknowledgments}}
\pano

We thank Oren Bergman, Tathagata Dasgupta, Eduardo Eyras, Peter
Goddard and Burt Totaro for useful discussions.

M.R.G.\ is grateful to the Royal Society for a University Research
Fellowship. S.S.-N.\ is supported by St John's College, Cambridge,
through a Jenkins Scholarship. This work is also partly supported by
EU contract HPRN-CT-2000-00122.   

\vskip1cm

\appendix{A}{K-theory analysis for the orbifold theory}

In this appendix we shall compute the groups \kpm; related
calculations have been done before in \refs{\gabstef,\BGK}. 
By a result of Hopkins (not published, but see \refs{\WittenK}),
$K_{\pm}$ can be related to the equivariant K-theory groups as  
\eqn\hop{
K_{\pm , cpt}(X)= K^{-1}_{\Zop_2 , cpt}(X \times \Rop^{1,0}) \,,}
where the suffix `$cpt$' denotes K-theory groups with compact support. 
Since $S^{2,0}$ is a retract of $S^{8-r}\times S^{2,0}$ we can write  
\eqn\kcpt{
K_{\pm} (S^{8-r} \times S^{2,0}, S^{2,0}) \, \oplus\,
K_{\pm, cpt}(S^{2,0})\,=\, 
K_{\pm, cpt} (S^{8-r}\times S^{2,0})\,.
}
Together with the relation \hop\ this implies
\eqn\spheres{
K_{\pm} (S^{8-r} \times S^{2,0}, S^{2,0})\, \oplus\, 
K_{\Zop_2, cpt}^{-1}(S^{2,0}\times \Rop^{1,0})\, =\, 
K_{\Zop_2, cpt}^{-1}( S^{8-r}\times S^{2,0}\times \Rop^{1,0})\,.
}
Further it follows from the K\"unneth formula \refs{\Atiyah} that 
\eqn\sakthree{
K^m_{\Zop_2, cpt} (X\times S^{2n})\, =\,
K^m_{\Zop_2, cpt}(X)^{\oplus 2}}
and
\eqn\sakfour{
K^m_{\Zop_2, cpt} (X\times S^{2n+1})\, =\,
K^m_{\Zop_2, cpt}(X) \oplus K^{m-1}_{\Zop_2, cpt}(X)\,.}
Putting all of this together we arrive at 
\eqn\kstepone{K_{\pm} (S^{8-r} \times S^{2,0}, S^{2,0}) = 
\left\{
\eqalign{K_{\Zop_2, cpt}(S^{2,0}\times \Rop^{1,0}) \qquad & 
\hbox{if $r$ is odd,} \cr 
K_{\Zop_2, cpt}^{-1}(S^{2,0}\times \Rop^{1,0}) \qquad & 
\hbox{if $r$ is even.}}\right.}
It therefore remains to compute the K-theory groups 
$K_{\Zop_2,cpt}^{*}(S^{2,0}\times\Rop^{1,0})$.\footnote{$^\dagger$}{We   
can also compute these by noting that the compactification of the
canonical line-bundle on $\Rop{\bf P}^1$, 
$S^{2,0}\times \Rop^{1,0}/\Zop_2$, is $\Rop{\bf P}^2$. The (complex)
K-theory groups for $\Rop{\bf P}^n$ have been computed in
\refs{\Adams,\Karoubi} and are given by 
$\widetilde{K}^m(\Rop{\bf P}^n)\, =\, \delta_{m, 2l}\Zop_{2^{[ n/2 ]}}$
where $l\in\Zop$. It therefore follows that 
$K_{\Zop_2,cpt}(S^{2,0}\times \Rop^{1,0})\, = \,
\widetilde{K}(\Rop{\bf P}^2)\,=\,\Zop_2$. Similarly, we obtain
$K_{\Zop_2,cpt}^{-1}(S^{2,0}\times \Rop^{1,0})\,=\,0$.}
These can be computed by using the long exact sequence
\eqn\les{ \eqalign{
\cdots\, \widetilde{K}^{-2}(S^{2,0}) & \,\arrow\,
K^{-1}_{\Zop_2,cpt}(S^{2,0}\times\Rop^{1,0} )  \,\arrow\,
K^{-1}_{\Zop_2}(S^{2,0})\, \arrow\,  K^{-1}( S^{2,0} )  \cr 
&\,\arrow\, K_{\Zop_2, cpt}(S^{2,0}\times \Rop^{1,0})
\,\arrow\,\widetilde{K}_{\Zop_2}(S^{2,0}) 
\,\arrow\, \widetilde{K}(S^{2,0})\,\arrow \cdots \,,}}
where $\widetilde{K}(X)$ denotes the reduced K-theory group, 
$K(X)= \widetilde{K}(X)\oplus \Zop$. Next we observe that
$S^{2,0}/\Zop_2=\Rop {\bf P}^1$, and therefore  
\eqn\proj{
K^n_{\Zop_2}(S^{2,0}) = K^n(\Rop{\bf P}^1)\,.}
Given the results of \refs{\Adams,\Karoubi} it
then follows that 
\eqn\proji{
\widetilde{K}^{2n}_{\Zop_2}(S^{2,0})= 0  \quad \hbox{and} \quad
K^{2n+1}_{\Zop_2}(S^{2,0}) =\Zop\,.}
This implies that $K^n(S^{2,0})=\Zop$ for all $n$ (since the reduced
and unreduced K-theory groups differ by $\Zop$ for $n$ even only).
Putting these results in the long exact sequence \les\ we then find 
\eqn\lesone{
0 \, \arrow  \, K^{-1}_{\Zop_2, cpt}(S^{2,0}\times \Rop^{1,0})
\, \arrow \, \Zop\, 
{\longrightarrow}^{\hskip-12pt \cdot 2\hskip6pt}\, 
\Zop \,\arrow\, 
K_{\Zop_2, cpt }(S^{2,0}\times\Rop^{1,0})\,\arrow\, 0 \,.}
The map between $K_{\Zop_2}^{-1}(S^{2,0})= K^{-1}(\Rop{\bf P}^1)$ and 
$K^{-1}(S^{2,0})$ is multiplication by 2 because the map between $S^1$
and $\Rop{\bf P}^1$ has degree 2.
It now follows from \lesone\ that 
\eqn\resone{
K_{\Zop_2, cpt}(S^{2,0}\times \Rop^{1,0})\cong \Zop_2 }
and
\eqn\restwo{
K^1_{\Zop_2, cpt}(S^{2,0}\times \Rop^{1,0})\cong 0 \,.}
Together with \kstepone\ we therefore arrive at the result
\eqn\Kresult{
K_{\pm}(S^{8-r} \times S^{2,0}, S^{2,0}) = \left\{
\eqalign{\;\;\Zop_2  &\qquad \hbox{if $r$ is odd,}\cr
         0\;\,      &\qquad \hbox{if $r$ is even.}}\right.}
This reproduces \KIIB\ and is in agreement with the D-brane spectrum
that was found by Gutperle in \refs{\gutperle}. 
\medskip

The K-theory groups characterizing the D$(r,s)$ branes for the
orbifold of the IIA theory by $(-1)^{F_L}\sigma_{\half}$ are 
\eqn\twoa{
K^{-1}_{\pm}(S^{8-r} \times S^{2,0}, S^{2,0}) \,.}
Using the identity that follows from \hop
\eqn\twoaiden{
K^{-1}_{\pm, cpt}(X)= K^{-1}_{\Zop_2, cpt}(X\times \Rop^{1,1})\,,}
and similar arguments as above, we obtain an analogue of \kstepone\
where now $\Rop^{1,0}$ is replaced by $\Rop^{1,1}$. We can then use
the suspension isomorphism 
\eqn\krel{
K^{-1}_{\Zop_2, cpt}(S^{n,0}\times \Rop^{1,1})
= K_{\Zop_2, cpt}(S^{n,0}\times \Rop^{1,0})}
and conclude that
\eqn\Kresulta{
K^{-1}_{\pm}(S^{8-r} \times S^{2,0}, S^{2,0}) = \left\{
\eqalign{\;\;\Zop_2  &\qquad \hbox{if $r$ is even,}\cr
         0\;\,      &\qquad \hbox{if $r$ is odd.}}\right.}
This reproduces \KIIA.

\listrefs

\bye